# ELECTRICAL DESIGN OF A CLEAN OFFSHORE HEAT AND POWER (CLEANOFF) HUB

| Maiken Borud Omtveit | Qian Long | Valentin Chabaud | Marte Ruud-Olsen | Steinar Halsne | Tor-Christian Ystgaard |
|---|---|---|---|---|---|
| ABB AS | ABB AS | SINTEF | ABB AS | ABB AS | ABB AS |
| Oslo | Oslo | Trondheim | Oslo | Oslo | Oslo |
| Norway | Norway | Norway | Norway | Norway | Norway |

*Abstract* - This paper presents an innovative offshore solution where oil & gas platform clusters are powered by a wind farm and a hydrogen hub. The results show a feasible off-grid design as an alternative to conventional electrification solutions.

To address the challenges of design and operation of such a system, a power system model of the equipment and control was developed in a power system simulator called Process Power Simulator (PPSim). Power fluctuations in the wind farm are modelled using a state-of-the-art method encompassing turbulence and wakes.

Various operation scenarios were used to evaluate the system design and find the right equipment size. An expensive component to over dimension is the battery energy storage system (BESS). The BESS power rating and energy capacity were found by running a combination of scenarios with extreme and natural wind variations, and contingencies.

The control strategy and ramp rates of electrolyzers have significant impact on both system performance and design. A ramp rate in the order of seconds as opposed to minutes will decrease the required BESS size by 60-70%. Choosing synchronized control of the electrolyzers can further reduce the BESS size by 15-20%. The simulations also revealed challenges to achieve self-sufficiency of hydrogen and potential design improvements are suggested.

## I. INTRODUCTION

In response to the need to combat climate change, the Norwegian government has initiated efforts to electrify the oil and gas installations on the Norwegian Continental Shelf. This paper addresses that with an energy hub concept that integrates offshore wind and hydrogen energy storage to supply the electric power consumption of oil and gas platforms. This concept can be an alternative for electrification of the Norwegian Continental Shelf when power from shore solutions are infeasible (e.g. due to too long distance from shore). Oil and gas platforms will be powered with renewable wind power. The excessive renewable power will be converted to hydrogen produced by electrolyzers at the hub while hydrogen will be used for generating electricity by co-located fuel cells during periods of wind power deficiency.

This concept is explored in a research project called Clean Offshore Heat and Power (CleanOFF) hub, a spinoff project from the LowEmission Research Centre supported by the Norwegian Research Council. The specific system topology and sizing of the CleanOFF hub has been developed in the CleanOFF project through previous studies, which the reader can find results from in [1,2,3].

In the CleanOFF hub concept, the system operates isolated from the grid, and conventional generators are decommissioned. At such low levels of rotating inertia in the system, introducing a large amount of variability from renewable wind, brings unique operational challenges. To ensure a reliable and stable operation, a BESS with grid-forming capability is required to stabilize frequency and voltage as well as to provide additional control functionalities. Dynamic simulations are conducted in this study to analyze the system behavior accounting for wind variations and system control dynamics.

This paper is organized as follows. Firstly, the method used to obtain the results is described, with details about the simulation tools, the scenario cases and how the model and control system was implemented. Then the results for short-term and long-term scenarios are presented, respectively. Finally, the conclusion presents the challenges and our recommendations for further studies on the CleanOFF system, considerations that may translate to offshore energy hubs in general.

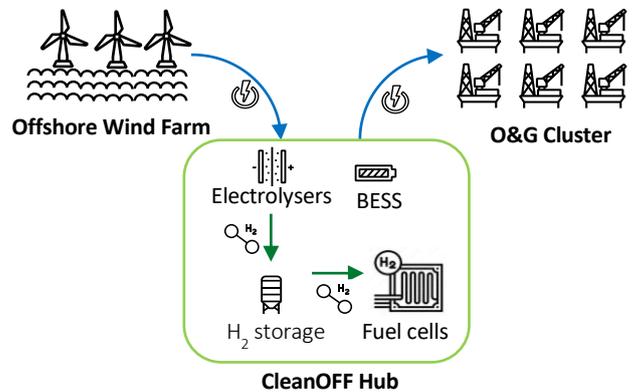

Fig. 1: CleanOFF hub concept

## II. METHOD

The framework of this study began with defining the base design of the system. Then, the base design and control with parameters from the CleanOFF system, as described in the following sections, was implemented in PPSim. An iterative approach was employed to study the system where simulations were run, and the results were evaluated to inform subsequent design improvements. Each iteration provided insights into the sensitivities of design variations in the system and informed the next set of design choices.

### A. Simulation Tool PPSim

PPSim is a sophisticated tool designed for real-time electrical simulation enabling comprehensive electrical system assessment and operator training [4]. By simulating steady-state and dynamic responses for voltages, currents, frequencies, and power flows in electrical systems, PPSim allows for the validation of control strategies, product solutions, procedures, and sequences in a safe and controlled setting. The simulator replicates the facility's electric and control systems and allows for studying both detailed dynamics on a short time scale and long-term behavior of the system. In this study, PPSim is used for early-phase testing of electrical topology and control concepts.

The PPSim models are based on electrical model standards and include parametrization for typical electrical components. Key features of the simulator include the ability to conduct full-scale testing of power control, load shedding, synchronization and other control functionalities provided by power management systems.

### B. System Topology and Modeling Approach

The electrical design of the CleanOFF hub system in this study is derived from the CleanOFF hub design featuring a GW-level wind farm. To balance computation efforts with model details, a system scaled down by a factor of 20 was chosen as the initial design. This approach allows for the inclusion of sufficient component model dynamics. It is assumed that the findings from the scaled-down system can be extrapolated to the original system, with the component capacities directly scaled up accordingly. However, the impacts of scaling on both short-term and long-term performance of the CleanOFF hub system should be investigated further.

The system is an islanded system without grid connection with the main bus at an offshore hub being rated at 66 kV nominal voltage. The BESS is the only one with grid-forming capability in the system. A screenshot of the PPSim model is displayed in Fig. 2. The following sections describe how the electrical components in the system were modelled.

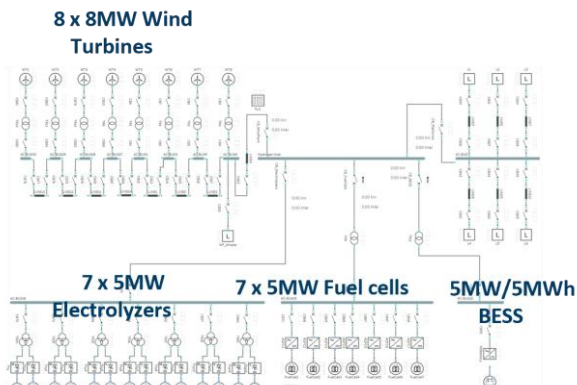

Fig. 2: SLD of initial CleanOFF system in PPSim

#### 1) Electrolyzer and Thyristor Rectifier Unit Model

Today, there are three main electrolyzer technologies on the market with different ramp rate characteristics. Alkaline electrolyzers is the most well-developed type and is the cheapest technology. The drawbacks are that it generally has slower ramp rate and has a large footprint, making it more unsuitable for offshore applications where space is limited. The proton exchange membrane (PEM) technology is a newer and more expensive technology. The advantage is that it has faster ramp rates and a smaller footprint. Lastly, there is the solid oxide electrolyzer, which is the newest commercially available technology. There is still little conclusive information about their ramp rates, but evidence suggests they are somewhere between the PEM and Alkaline [5].

In this study's electrolyzer model, the electrical domain dynamics are represented by the I/V curve and ohmic losses. The nominal production is 1000 $Nm^3$/h for each 5MW unit, and full operating range is assumed, from 0-100% load. Different electrolyzer technologies on the market are represented through the ramp rate of the electrolyzer. The fast ramp rate is 11s from zero to full load, whereas the slow electrolyzer ramp rate is 706s, equivalent to almost 12min.

Each pair of two 2.5 MW electrolyzers are connected to the electrical network through a 12-pulse thyristor bridge in a master/slave configuration, a three-winding transformer and switchgear. Each three-winding transformer and downstream units are denoted as one train for this study. The electrolyzer plant consists of seven 5 MW electrolyzer trains.

#### 2) Fuel Cell and Active Front End Converter Model

The fuel cell plant consists of one 35 MVA 66 kV/400 V transformer, one 400 V switchgear and seven 5 MW/5 MVA fuel cell blocks. Each fuel cell block is interfaced to the AC grid with one Active Front End (AFE) converter of the same rating as the fuel cell.

The chosen fuel cell stack is based on a PEM fuel cell stack data sheet from Nedstack [6]. The ramp rate of the fuel cell is assumed to be 9%/sec (10 second from zero to full load), which is notably as fast as the fast-ramping electrolyzers in the model. Each fuel cell consists of 12 stacks in series and approximately 31 stacks in parallel.

The fuel consumption is derived from the data sheet stoichiometry, fuel and oxidant composition, and the current/voltage characteristic. For this project a constant fuel stoichiometry is chosen to be 1.25. The constant stoichiometry ratio for air flow is assumed to be 2. The fuel is assumed to consist of 100% pure $H_2$, and the oxidant is assumed to be air consisting of 21% $O_2$.

#### 3) BESS Model

The BESS model consists of one transformer, one four-quadrant converter and one battery storage. The battery storage unit is a simplified representation of multiple DC battery stacks. The BESS converter acts as a grid forming source that actively regulates the CleanOFF hub system frequency and voltage and balances any mismatch at any instant for active and reactive power. The BESS converter is connected to the main 66 kV bus via a 20 MVA 66 kV/480 V transformer.

The grid-forming control in this study has been implemented based on a specific converter made in-house. It applies the method of virtual synchronous machine (VSM) to converter control to provide inertia and primary droop response to the system. It is critical to ensure that BESS dynamics have been validated as this will be the dominant factor driving the MW requirements of BESS. Therefore, the grid-forming control in PPSim has been validated against the one in PowerFactory via dynamic simulation. Fig. 3 shows voltage and frequency dynamics of the grid-forming BESS converter after grid disconnection event occurs at t = 30 sec, indicating that the PPSim BESS converter model has good consistency with the PowerFactory model.

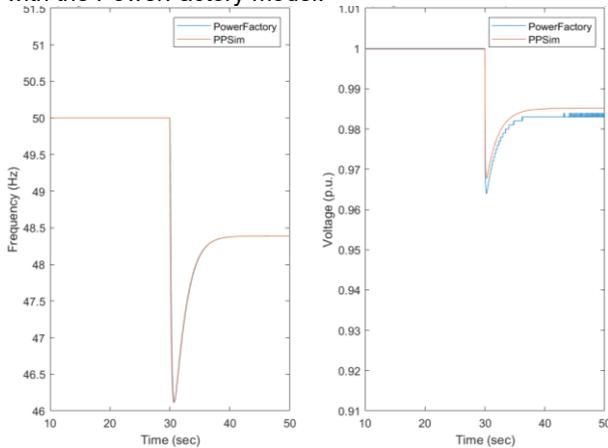

Fig. 3: Comparison of frequency and voltage dynamics of grid forming BESS between PPSim and PowerFactory

#### 4) Platform Model

There are six offshore platforms with a total load consumption of 27.5 MW, modelled as static loads, as platforms tend to operate at constant loading. Two of the platforms are at 5.75MW and four are at 4MW. It is assumed that all the platforms operate at a constant power factor of 0.8 with constant power load characteristics.

#### 5) Wind Farm Model

The wind park is represented by 8 individual type 4 wind turbines rated at 8 MW. The wind turbine model is based on IEC standard 61400-27-1, including turbine controller and pitch controller. Each wind turbine has a cut-in wind speed of 4 m/s, and a cut-out speed of 25 m/s.

The layout of the wind farm is presented in Figure 2. The wind turbines are arranged in two rows, each facing upwind at all times, with a spacing of 5 diameters between them.

With regards to wind speed variability and rapid power fluctuations, this setup represents a conservative assumption. Due to the smoothing effect of a wind farm, a larger wind farm spanning over a greater geographical area may exhibit less power fluctuations for the same wind speed series.

The wind farm is regulated by a power plant controller modelled based on IEC standard 61400-27-1. The plant controller operates the wind farm at unity power factor and with closed loop control. It distributes active power setpoints to the individual wind turbines. The wind turbines have been implemented to accept active power limit signals from the plant controller. Note that this is not the state-of-practice in operational wind farms today, where active power control is rudimentary (open-loop, no power tracking feature, uniformly distributed across turbines). Advanced active power control considering i.e. wake advection delay to temporarily increase power has not been considered. For the scenarios run in this study, the voltage on the wind farm main bus stayed within acceptable limits.

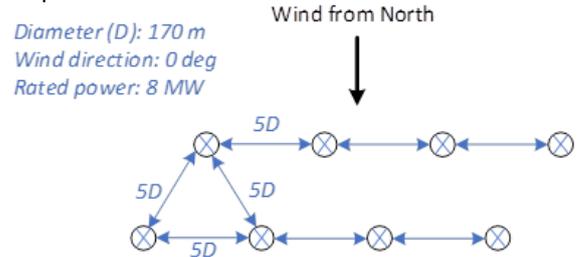

Fig. 4: Wind farm physical layout of the 8 wind turbines

### C. Control System Philosophy

The CleanOFF hub system is designed to efficiently regulate both frequency and voltage, to ensure operational stability. The control integrates the BESS, wind farm, electrolyzer plant and fuel cell plant to dynamically balance the system. The power management ensures nominal frequency regulation, state-of-charge (SOC) control of the BESS, and effective energy conversion to hydrogen. Additionally, the BESS and AFE converters support reactive power compensation. The following sections outline the methodologies used in active power and frequency control and reactive power and voltage regulation.

#### 1) Active Power and Frequency Control:

The high-level active power and frequency control philosophy of the CleanOFF hub system is as follows:
- Power management system always monitors the system frequency and ensures that the system frequency is regulated at nominal.
- Power management system always monitors the SOC of BESS and ensures that the SOC is regulated at the preferred target.
- Power management system sends active power setpoint to plant controllers of the wind farm, the electrolyzer plant and the fuel cell plant.
- BESS is charged when SOC is lower than the preferred target and is discharged when SOC is higher than the preferred target.
- All electrolyzers are controlled in sequence by default, meaning that electrolyzers will start up or shut down one by one. The electrolyzer plant controller can also be configured to operate in synchronization, meaning that electrolyzers will start up or shut down simultaneously.
- All fuel cells are controlled in sequence by default, meaning that fuel cells will start up or shut down one by one.
- When there is excess power, fuel cells in operation are transition to standby and electrolyzer trains are brought online in order to convert renewable power to hydrogen production.
- When electrolyzers run at full load level and there is still power surplus, the wind farm is at curtailment mode and is set to an upper limit.
- When there is power deficiency, electrolyzer trains are switched to standby mode and fuel cells are brought online to supply power by converting stored hydrogen.
- When there is no available wind power, fuel cells, together with BESS, supplies fully oil and gas

(O&G) platform loads, given there is hydrogen stored in the tank.
- If there is a short supply of hydrogen, fuel cell needs to ramp down and load shedding scheme is triggered to disconnect low-priority loads until the system enters a full shutdown.

*2) Reactive Power and Voltage Control*

The reactive power consumption in the system mainly comes from two sources: the reactive power demand from O&G platform clusters and reactive power consumed by the thyristor rectifier units. The former is considered constant while the latter varies with the loading levels of electrolyzers.

Reactive power compensation in this system is provided by a combination of the BESS converter and the AFE converters from the fuel cell plant. The BESS converter regulates the voltage at its terminal and provides reactive power for any instantaneous reactive power imbalance in the system. The AFE converters compensate reactive power in such a way that the BESS converter contributes with as little reactive power as possible, and provide nominal voltage to the system.

In the CleanOFF hub system, as electrolyzers and fuel cells don't operate in parallel at any instant, AFE converters have the capacity to support reactive power demand from thyristor rectifiers the majority of the time. During low wind condition, fuel cells need to supply active power to O&G platforms, and there will be no free capacity from AFE converters to provide reactive power. In this case, the BESS converter will provide reactive power instead. The wind farm runs in unity power factor and doesn't participate in reactive power control, although it could be, and is considered in further work.

*D. Objective of Simulation Results and Introduction of cases*

The overall objective of this study is to conduct a concept feasibility study for the CleanOFF hub system. The purpose of this study is to 1) evaluate the power system control philosophy of the CleanOFF hub addressing electrical stability challenges; 2) evaluate and recommend improvements to the electrical design of the CleanOFF hub system, including topology, component rating and technology selection.

The following sections present the short-term cases and long-term cases. Short-term studies capture the most challenging operating conditions, intended to examine the MW rating of the BESS while the objective of long-term studies is to examine the MWh rating of the BESS, the capacity of the electrolyzers and fuel cells, the hydrogen storage tank size, etc.

*1) Short-Term Cases*

Short-term scenarios are used to validate that the system can handle severe cases of natural wind variations, and contingencies of load and generation trip. Two types of design variations are implemented in the short-term cases, as they are the driving factor for the BESS MW requirement:
- **Ramp rate of electrolyzers:** typical values of ramp rates are applied based on the publicly available datasheets from main electrolyzer manufacturers, where 11 sec from 0% to 100% is used to represent fast-ramping electrolyzers and 706 sec from 0% to 100% is used to represent slow-ramping electrolyzers.
- **Operational strategy of electrolyzer plant:** two strategies are implemented to operate the electrolyzer plants. One is operating in sequence while the other operating in synchronization. In sequence refers to electrolyzers starting up and shutting down one by one while in synchronization refers to all the electrolyzers always operating at the same loading level.

Additionally, two N-1 contingency cases are considered, where one critical component disconnects at the most critical time instant. The most critical instant is identified in the simulation as the instant when the BESS is at its highest charging or discharging rate. This is the time when it is most likely that the BESS reaches its current limit, and that system instability occurs with a downsized BESS. Two contingency cases are applied and include:
- **Trip of the biggest load**: this corresponds to the trip of the biggest platform load while the BESS is charged at its highest rate.
- **Trip of the biggest generation unit**: this corresponds to one 8 MW wind turbine (WT) trip while the BESS is discharged at its highest rate.

Table I summarizes the short-term scenarios that were run for varying electrolyzer (ELY) ramp rates, control strategies and events.

TABLE I
SHORT-TERM SCENARIOS

| Study Case | Wind Scenario | ELY Ramp Rate | ELY Control Strategy | Event |
|---|---|---|---|---|
| S1 | Severe variations around 10 m/s | 706 s from zero to full load | In sequence | N/A |
| S2 | Severe variations around 10 m/s | 706 s from zero to full load | In synch | N/A |
| S3 | Severe variations around 10 m/s | 11 s from zero to full load | In sequence | N/A |
| S4 | Severe variations around 10 m/s | 706 s from zero to full load | In sequence | WT trip |
| S5 | Severe variations around 10 m/s | 11 s from zero to full load | In sequence | WT trip |
| S6 | Severe variations around 10 m/s | 706 s from zero to full load | In sequence | Load trip |
| S7 | Severe variations around 10 m/s | 11 s from zero to full load | In sequence | Load trip |

*2) Long-Term Cases*

The long-term cases are used to evaluate the system's long-term performance. Key performance metrics such as plant utilization, capacity factor and longest shutdown period are analyzed. These metrics provide insights into the validity of the system design and suggest improvements to maximize efficiency and reliability while minimizing the investment costs. Our initial evaluation indicates that a 35 MW fuel cell plant is oversized for the system, resulting in zero utilization of the fuel cell block #7. Therefore, we propose two design variations with smaller fuel cell plant size, as displayed in Table II.

The main difference between Design 1 and Design 2 is that Design 2 has one less fuel cell unit. Given that the peak load for the O&G platforms is approximately 27 MW compared to 25 MW fuel cell plant, the BESS must continue discharging to cover the gap between load consumption and fuel cell power output when wind power is unavailable. Therefore, a larger BESS energy capacity is required in Design 2. The design variations are assessed in section III.C below.

TABLE II
DESIGN PARAMETERS FOR THE CLEANOFF HUB SYSTEM

| Design | Initial Design | Design 1 | Design 2 |
|---|---|---|---|
| Wind farm | 64 MW | 64 MW | 64 MW |
| Electrolyzer plant | 35 MW | 35 MW | 35 MW |
| Fuel cell plant | 35 MW | 30 MW | 25 MW |
| BESS | 10 MW/ 10 MWh | 10 MW/ 10 MWh | 10 MW/ 300 MWh |

## III. RESULTS

This section presents dynamic simulation results that are used to determine the capacity requirements for the CleanOFF hub. Firstly, the simulated wind timeseries used in both short-term and long-term studies are presented. Then, the results from the short-term and long-term simulations are presented which determine the capacity requirements for the CleanOFF hub. The capacity requirements include, but are not limited to, MW and MWh capacity of BESS, MW capacity of electrolyzer plants, MW capacity of fuel cell plants, hydrogen storage tank capacity.

### A. Wind Timeseries

Power fluctuations from the wind farm are an essential input to the system, as they determine ramp rates and energy fluctuations that the BESS, fuel cells and electrolyzers must cope with. Knowledge about the individual contributions from each turbine is also valuable for local effects (e.g. control of power electronics). To satisfy these requirements, a state-of-the-art framework capable of modelling wind fluctuations in offshore wind farms with power fluctuations in mind has been used. The approach is similar to that used in [3] to which the reader is redirected for a more detailed description. In short, care is taken to capture the correlation in wind speed between turbines caused by large, low-frequency coherent flow patterns (i.e. turbulent vortices) and wakes. To this end, the dedicated tools FLAggTurb [7] and DIWA [8] developed at SINTEF have been used. Turbulence being stochastic, multiple realizations of timeseries should be used in Monte Carlo simulations. However, consistently with the current purpose of demonstration and not targeting extreme conditions, only one realization is presented in this paper for simplicity.

*1) Short-Term 2-hours Wind Speed Timeseries*

Consistently with the wind turbine model (from the IEC standard 61400-27-1) using quasi-steady power curves, timeseries of rotor-averaged wind speed for each turbine are generated. They represent severe yet likely (i.e. not extreme, very rare events) operating conditions where power fluctuations mostly arise from turbulence in the ambient wind. The most challenging conditions are selected for the short-term studies. They correspond to the wind turbines' rated wind speed (10 m/s as shown in Table I), i.e. just before the turbine controller stabilizes power to its rated value by pitching blades; and to a high (but not extreme) turbulence intensity. The 2-hours simulated wind speed timeseries for each turbine is shorn in Fig. 5.

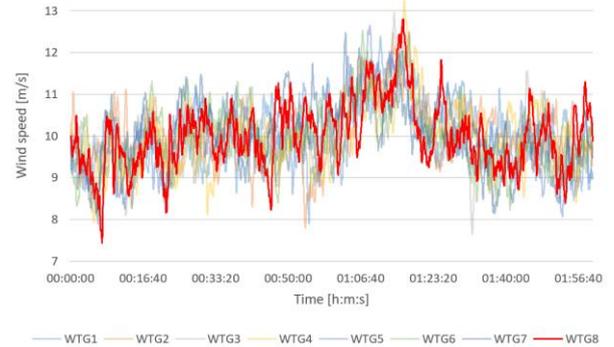

Fig. 5: 2-hours wind speed timeseries for each of the eight wind turbines in the wind farm

*2) Long-Term Power Timeseries*

For long-term studies, wind speed has been directly translated into power using quasi-steady power curves. 2-hours short-term timeseries matching mean conditions obtained from the NORA-3 weather reanalysis database over one year are then assembled, following the procedure described in [9]. Fig. 6 displays the timeseries for available power of the wind farm during one year.

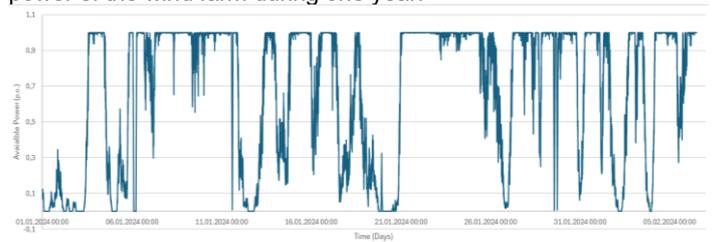

Fig. 6: Available power timeseries from wind farm for one year

### B. Short-Term Simulation Results

The simulation results for the selected short-term cases, also considered as "boundary cases", are presented in this section. These short-term cases are used to evaluate the system's dynamic responses and stability for the most severe wind conditions and selected contingencies.

*1) Normal Operation under Variable Wind Conditions*

This section presents the results from simulating the 2-hour window with the most rapid changes in wind speed. Wind variation introduces active power imbalance to the system, which in turn causes frequency deviation. Secondary frequency control manages over-frequency by ramping up electrolyzers and curtailing wind power, and under-frequency by ramping up fuel cells. Active power mismatches beyond the control bandwidth of secondary frequency control are managed by the grid-forming control of the BESS.

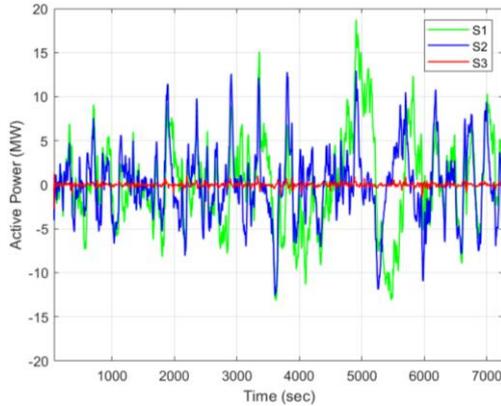

Fig. 7 BESS active power output for S1 (slow-ramping electrolyzers running in sequence), S2 (slow-ramping electrolyzers running in synchronization) and S3 (fast-ramping electrolyzers running in sequence)

The BESS charging and discharging characteristics for the 2-hour period are shown in Fig. 7. In Scenario 1 (S1), where electrolyzers with slow ramp rates operate sequentially, the BESS must be charged and discharged at a high rate to mitigate the active power imbalance introduced by the wind variation. In Scenario 2 (S2), where electrolyzers with slow ramp rate run in synchronous operation, the BESS still exhibits relatively high charging and discharging rates, but these rates are lower compared to S1. This is consistent with the fact that electrolyzers running in synchronization provide a higher ramp rate for active power regulation at the plant level compared to electrolyzers running in sequence. As can be seen from the red curve, the BESS output power which is required for active power balance is limited in Scenario 3 (S3), where fast-ramping electrolyzers are operated in sequence.

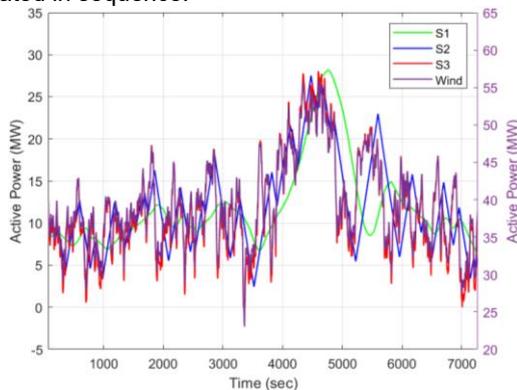

Fig. 8: Electrolyzer plant active power consumption for S1 (slow-ramping electrolyzers running in sequence), S2 (slow-ramping electrolyzers running in synchronization) and S3 (fast-ramping electrolyzers running in sequence)

Fig. 8 illustrates the active power demand of the electrolyzer plant, with the active power generation of the wind power plant presented on the secondary y-axis on the right. In S3, fast-ramping electrolyzers closely track wind variation. In S2, slow-ramping electrolyzers operating synchronously track wind variation to some extent but deviate when the wind ramps up and down faster than the electrolyzer plant's capability. In S1, slow-ramping electrolyzers running sequentially face significant difficulties in following wind variation.

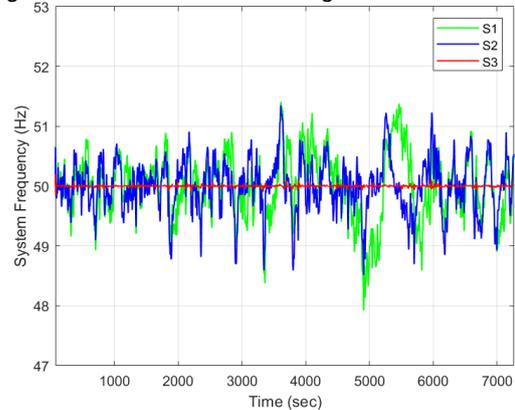

Fig. 9: System frequency for S1 (slow-ramping electrolyzers running in sequence), S2 (slow-ramping electrolyzers running in synchronization) and S3 (fast-ramping electrolyzers running in sequence)

Fig. 9 shows the system frequency at the main bus of the CleanOFF hub. There are ±2 Hz (±0.040 p.u.) and ±1.3 Hz (±0.026 p.u.) deviations for S1 and S2, respectively. Frequency is maintained at nominal 50 Hz within ±0.2 Hz (±0.004 p.u.) deviation in S3. Maximum charging and discharging rates of BESS for all three scenarios are presented in Table III. In S1, a BESS larger than 18.8 MW is required to handle wind variation during normal operation. S2 necessitates a BESS no lower than 13.0 MW, while S3 only requires a BESS with at least 1.5 MW capacity. Fast-ramping electrolyzers operating synchronously are not investigated in this study, as running them in sequence has already demonstrated sufficiently fast ramp rates for closely tracking wind variation.

TABLE III
MAXIMUM CHARGING AND DISCHARGING RATE OF BESS DURING MOST SEVERE WIND CONDITIONS

| Study Case Number | Max Charging Rate | Max Discharging Rate |
|---|---|---|
| CleanOFF-S1 | 13.1 MW | 18.8 MW |
| CleanOFF-S2 | 12.6 MW | 13.0 MW |
| CleanOFF-S3 | 1.5 MW | 1.2 MW |

The voltage at the main bus of the CleanOFF hub is shown in Fig. 10. In all scenarios, the voltage is maintained around the nominal value with ±0.015 p.u. deviation. Fig. 11 shows the BESS reactive power output, indicating that the BESS reactive power output is independent of its active power output and electrolyzer ramp rates. Thanks to the reactive power control implemented at the fuel cell plant, reactive power contribution from the BESS is negligible. The reactive power control effectively helps minimize the size of battery storage converter. As seen in Fig. 11, the fuel cell converters, in all the scenarios, manage to compensate variable reactive power demand from electrolyzers that have varying loading levels.

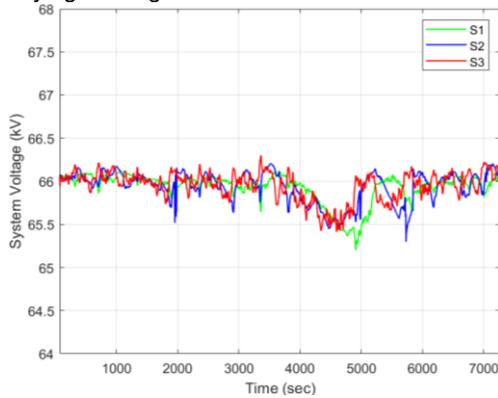

Fig. 10: System voltage for S1 (slow-ramping electrolyzers running in sequence), S2 (slow-ramping electrolyzers running in synchronization) and S3 (fast-ramping electrolyzers running in sequence)

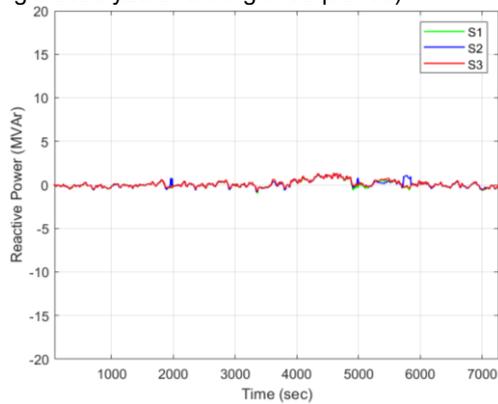

Fig. 11: BESS reactive power for S1 (slow-ramping electrolyzers running in sequence), S2 (slow-ramping electrolyzers running in synchronization) and S3 (fast-ramping electrolyzers running in sequence)

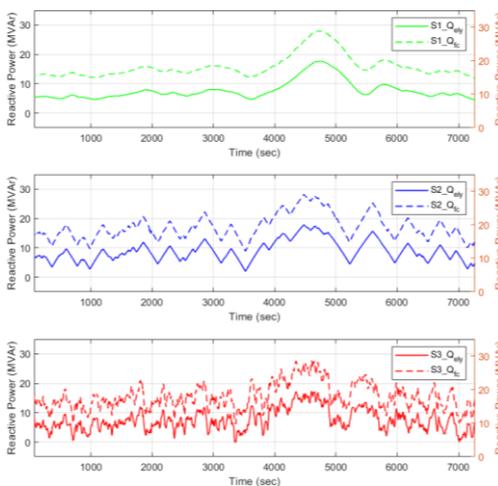

Fig. 12: Reactive power of electrolyzers and fuel cells in S1, S2 and S3

### 2) N-1 Event under Variable Wind Conditions

This section presents the results from simulating the contingency events. The wind variations are the same as in the previous cases. The contingency events are triggered at the worst possible time instant during the 2-hour window. For example, a wind turbine is tripped while the BESS is being discharged the most to balance active power. These contingency events further stress the BESS, requiring an instantaneous active power regulation. This regulation is too fast to be provided by secondary frequency control and must be provided by grid forming control of the BESS. It is worth noting that other N-1 scenarios, such as failure on the incoming cable of wind power plant or O&G platforms, are not included in this study but should be considered when such an offshore hub is designed. How to handle such contingencies will be considered in further work.

Fig. 13 shows the active power output of the BESS in Scenario 4 (S4). In S4 the system has slow ramping electrolyzers running in sequence, meaning there is less contribution to frequency support from the electrolyzers compared to in the other cases. A wind turbine trips at the worst time, at t = 6710 sec, when the BESS is being discharged at 18.8 MW active power, as seen in Table III. It can be seen that the peak discharging rate of BESS is 23.5 MW.

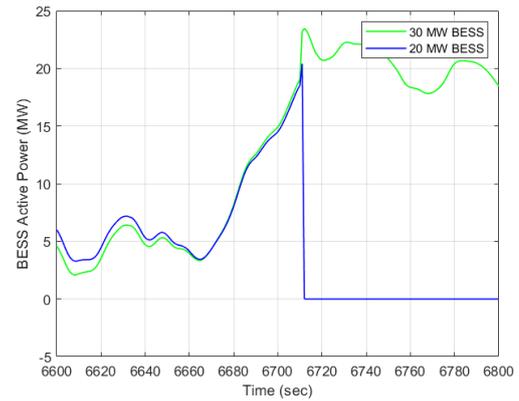

Fig. 13: BESS active power with the wind turbine trip event for S4

Fig. 14 shows the active power output of the BESS in Scenario 6 (S6), when there is slow ramping electrolyzers running in sequence and a load trips. At t = 5430 sec, the largest platform load trip occurs at the worst time, when the BESS is charging at 13.1 MW, as indicated in Table III. It can be seen that the peak charging rate of BESS is 18.1 MW.

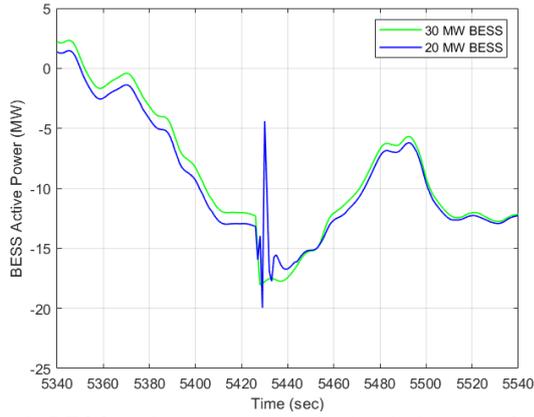

Fig. 14: BESS active power with the load trip event for S6

Fig. 15 shows the active power output of the BESS in Scenario 5 (S5). In S5 the system has fast ramping electrolyzers running in sequence, and a wind turbine is tripped at the worst possible time, when the BESS is being discharged at a maximum. The wind turbine trips at t = 2190 sec when the BESS is being discharged at 1.2 MW active power, as indicated in Table III. It can be seen that the peak discharging rate of BESS is 5.6 MW.

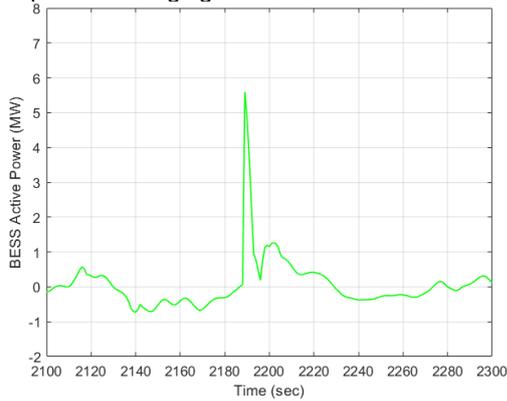

Fig. 15: BESS active power with the wind turbine trip event for S5 for 10 MW BESS

Fig. 16 shows the active power output of the BESS in Scenario 7 (S7), when there is fast ramping electrolyzers running in sequence and a load trips. At t = 5960 sec, the largest platform load trip occurs at the worst time when the BESS is charging at 1.5 MW, as indicated in Table III. It can be seen that the peak charging rate of the BESS is 5 MW.

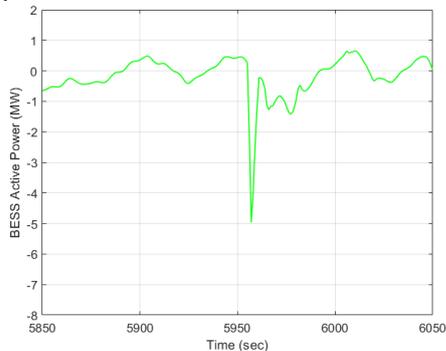

Fig. 16: BESS active power with the load trip event for S7 for 10 MW BESS

The findings above suggest that additional MW capacity for the BESS should be considered to maintain system operational during contingencies. Considering the results from S4 and S6, it is evident that quite a large BESS is required to maintain stability with slow ramping electrolyzers running in sequential operation. Considering the results from S5 and S7, it is evident that a much smaller BESS is required to maintain stability with fast ramping electrolyzers running in sequential operation. Considering that SOC control may engage the BESS in either charging or discharging mode, a certain headroom capacity should be reserved. Based on the simulation results it is found that a 30 MW BESS is required when utilizing slow-ramping electrolyzers operating sequentially and a 10 MW BESS when utilizing fast-ramping electrolyzers operating sequentially, for the cases considered in this study. It is worth noting that to increase system reliability, a larger BESS is always desired. It might be worthwhile to conduct a contingency analysis in further work to explore the relationship between reliability improvements and investment in BESS MW capacity.

### C. Long-Term Simulation Results

The simulation results for the selected long-term cases are presented in this section. Year-long performance metrics for Design 1 are presented in Table IV. The wind power plant (WPP) exhibits a high utilization rate of 0.91. This means the wind farm produces power 91% of the time during the year. The longest shutdown period it experiences is of nearly one and a half days, when there is no or very low wind. It also demonstrates a high capacity factor of 0.51, which aligns well with the capacity factors of 0.4 to 0.5 observed in new offshore wind farms [10]. The year-long curtailed energy is approximately 7GWh, equivalent to 2.4% of the wind farm annual energy wind generation, indicating efficient utilization of the wind farm. The electrolyzer plant has a utilization rate of 0.53, with its longest shutdown lasting over a week. The fuel cell plant (FC) has a utilization rate of 0.47 and the longest shutdown period of five and a half days. As the fuel cell plant serves as the back-up capacity, its capacity factor is 0.30, which is relatively low.

TABLE IV
KEY YEAR-LONG METRICS FOR DESIGN 1

| Plant Name | Utilization Rate | Longest Shutdown Period (days) | Curtailed Energy (GWh) | Capacity Factor | Load Factor |
|---|---|---|---|---|---|
| WPP | 0.91 | 1.44 | 6.98 | 0.51 | N/A |
| ELY | 0.53 | 7.36 | N/A | N/A | 0.41 |
| FC | 0.47 | 5.49 | N/A | 0.30 | N/A |

The year-long performance metrics for Design 2 are presented in Table V. The wind farm's long-term performance is very similar to that of Design 1. Compared to Design 1, the year-long curtailed wind energy is slightly reduced by 0.05 GWh, which is utilized for charging BESS to its target SOC. The electrolyzer plant and the fuel cell plant also show a comparable performance to Design 1, with a marginal increase in the longest shutdown period for the electrolyzer plant and a slight decrease for the fuel cell plant. As clearly illustrated in Fig. 17 and Fig. 18, Design 2 demonstrates a slightly lower utilization rate of the electrolyzer trains but a marginally higher utilization rate of the fuel cell units. As the fuel cell plant rated capacity is reduced in Design 2, the capacity factor increases up to 0.35.

TABLE V
KEY YEAR-LONG METRICS FOR DESIGN 2

| Plant Name | Utilization Rate | Longest Shutdown Period (days) | Curtailed Energy (GWh) | Capacity Factor | Load Factor |
|---|---|---|---|---|---|
| WPP | 0.91 | 1.45 | 6.93 | 0.51 | N/A |
| ELY | 0.53 | 7.40 | N/A | N/A | 0.42 |
| FC | 0.47 | 5.48 | N/A | 0.35 | N/A |

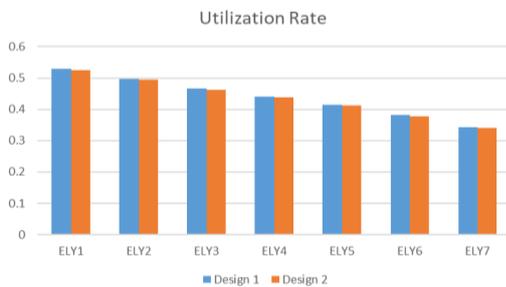

Fig. 17: Utilization of each electrolyzer train for both designs

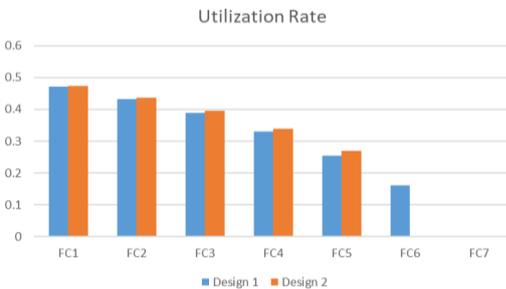

Fig. 18: Utilization of each fuel cell unit for both designs

The curtailed wind generation in these two designs above occurs primarily because the wind farm does not reach its full capacity when the electrolyzer plant operates at full loading level. To better utilize the curtailed wind energy, the electrolyzer plant could operate at overloading level to match the wind farm rated capacity, thereby increasing hydrogen production. However, the potential improvement is limited.

The comparison above shows that Design 2 marginally improves the long-term performance of the wind farm and the fuel cell plant while slightly reducing the utilization of the electrolyzer trains. This shift is primarily due to the increased involvement of the BESS in supplying the load during no wind or low wind periods in Design 2. In this case, the BESS needs to be charged by the wind farm and the fuel cell to reach its target SOC, when there is sufficient wind.

Fig. 19 illustrates the BESS SOC for both Design 1 and Design 2. As seen in Design 2, there are multiple instances where the BESS SOC drops to a lower range and then recovers to the target SOC. The BESS SOC is kept within a tight band around the target value in Design 1. The result proves the effectiveness of the BESS SOC control design and demonstrates a proper selection of MWh capacity for both designs. It is worth noting that Design 2 ends with a very large BESS MWh capacity. It indicates that the MWh requirement of the BESS is sensitive to the fuel cell plant capacity. If the fuel cell plant can fully meet the demand from O&G platforms, the BESS MWh capacity requirements will be significantly reduced.

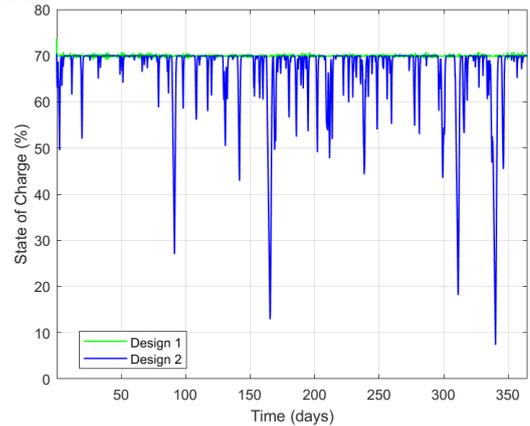

Fig. 19: State of charge for both system designs.

### D. Self-Sufficiency of Hydrogen

As shown in Fig. 20, the year-long simulation demonstrates that neither Design 1 nor Design 2 achieves a fully self-sufficient system, as both exhibit a net consumption of hydrogen over the course of the year. Design 1 requires a net hydrogen import of 2,433 ton while Design 2 requires import of 2,415-ton hydrogen. This comparison shows that reducing fuel cell capacity and increasing BESS energy capacity for load backup has limited impact on achieving self-sufficiency of hydrogen in the CleanOFF hub system, considering the CleanOFF hub system is intended to operate in an off-grid manner. Below are the recommended directions for the system design to achieve self-sufficiency of hydrogen.

- **Increase the capacity of both wind farm and electrolyzer plant:** Given the efficient utilization of the wind farm with the existing capacity, simply increasing the capacity of the wind farm would result in more curtailed wind generation, while increasing only the electrolyzer capacity would decrease the utilization of electrolyzer trains. In addition, considering the need for net hydrogen imports throughout the year, it is evident that hydrogen production must be increased by operating more electrolyzers. The additional load demand from these electrolyzers should be met by additional wind generation. Therefore, a CleanOFF hub design with larger wind farm and electrolyzer capacity should be studied to determine the optimal number of wind turbines and electrolyzers required to achieve hydrogen consumption neutrality over an extended period. Design 3 has been introduced to study the upscaling factor for the electrolyzer plant and the wind farm, where the electrolyzer plant has been increased by a factor of 1.6, 1.8 and 2.0, respectively, and the wind farm has been increased to match the electrolyzer plant peak demand. In Fig. 20 it is seen that a factor of no less than 1.8 needs to be applied to the capacity of the electrolyzer plant to achieve hydrogen neutrality at the end of the year, and it leads to 88 MW and 63 MW capacity for the wind farm and the electrolyzer plant, respectively. As to what the minimum size of hydrogen storage is to achieve self-sufficiency, it needs to be further investigated.

- **Integrate backup generators with fuel cell:** To achieve a self-sufficient system design without relying on hydrogen imports, adding back-up generators (either at the hub or on the O&G platforms) will significantly reduce hydrogen consumption and lead to a more reasonable size for hydrogen storage tank. Additional control strategies for both fuel cells and electrolyzers will likely be necessary such that the stored hydrogen level is effectively regulated.

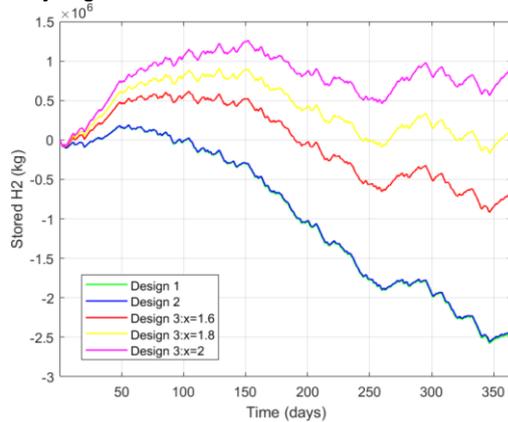

Fig. 20: Net stored hydrogen in kg for both designs and designs with larger wind farm and electrolyzer capacity

## IV. CONCLUSION

The electrification of an offshore platform cluster by means of a wind farm and a hydrogen hub has been validated and studied. The effect of varying operational strategy and ramp rate of the electrolyzers has been studied with simulated scenarios in PPSim. Simulated wind series for both short-term and long-term scenarios were obtained, and contingencies were imposed on the system in short-term scenarios to evaluate the N-1 stability of the system, giving insights into the minimum size of the BESS MW capacity. The iterative method used in this study of changing design parameters informed by running simulations allowed for optimizing parameters during the long-term simulations and proved efficient in gaining insights to the relative sizes of the components in the system.

To achieve an effective design for an offshore hydrogen- and wind-powered hub aimed at decarbonizing O&G assets, this report proposes the following key modifications to the initial design:

- **Fuel Cell Plant Capacity:** The original system has oversized the fuel cell plant. It is recommended to reduce its capacity to better match the peak demand of the O&G platform cluster.
- **Electrolyzer Ramp Rates**: A 100 MW grid-forming BESS is feasible only for GW-level offshore energy hub if the electrolyzer plant consists of fast-ramping electrolyzers capable of reaching full loading within tens of seconds. If slower-ramping electrolyzers are used, a grid-forming BESS with a larger MW rating is expected.
- **Hydrogen Management:** The system currently requires a net hydrogen import, making the usage of the hydrogen storage tank impractical. To eliminate the need of hydrogen import by ship, two alternatives design approaches are suggested: either increase the capacity of both the WF and the electrolyzer plant, or replace part of the fuel cell plant capacity with generators for load backup.
- **Reactive Power Control:** Given the assumption that thyristor rectifiers are used for electrolyzers and AFE rectifiers are used for fuel cells in this study, a reactive power control strategy is necessary. One potential approach, without adding extra reactive power compensating units, is to leverage the reactive power support capability of the AFE rectifiers at the fuel cell plant while minimizing the reactive power output from the BESS.

## V. FURTHER WORK

Through the study, several main directions for further work have been identified. System-level design enhancements should focus on eliminating the need for hydrogen imports. Additionally, further studies should aim to improve the modeling of wind farms and hydrogen plants, particularly in enhancing the representation of short-term dynamics of wind variation and hydrogen production processes in electrical simulations. These aspects are crucial as they significantly influence the BESS size. Exploring the integration of reactive power capabilities of wind turbine converters into reactive power control is another important area. Lastly, introducing demand flexibility in O&G platforms could reduce the required BESS size, thereby improving the project's feasibility from a cost perspective.

## VI. ACKNOWLEDGEMENTS

This research is conducted with ABB in-kind contribution to the research centers LowEmission (grant number 296207), NorthWind (grant number 321954), and Hydrogeni (grant number 333118), which are supported by the Norwegian Research Council.

## VII. REFERENCES

[1] Damaceno, D. S., Treider, T. & Svendsen, H. (2024, August). Wind Turbines and Hydrogen-Based Energy Storage Hub Concept for Offshore Oil and Gas Platforms in the Norwegian Continental Shelf. In *Proceedings of the International Conference on Ocean, Offshore and Arctic Engineering* (Vol. 7, p. V007t09A102). American Society of Mechanical Engineers.

[2] Dos Santos Mota, D., Fernando Alves, E., Sanchez-Acevedo, S., Svendsen, H. & Tedeschi, E. (2022, June). Offshore wind farms and isolated oil and gas platforms: Perspectives and possibilities. In *Proceedings of the International Conference on Offshore Mechanics and Arctic Engineering* (Vol. 10, p. V010T11A048). American Society of Mechanical Engineers.

[3] Dos Santos Mota, D., Haugdal, H., & Chabaud, V. (2024, November). Analysing a grid-forming storage hub for an offshore platform cluster supplied by wind energy. In *Journal of Physics: Conference Series* (Vol. 2875, No. 1, p. 012008). IOP Publishing.

[4] ABB. "Process Power Simulator". https://new.abb.com/oil-and-gas/products/automation/process-power-simulator

[5] Martinez Lopes, V. A., Ziar, H., Haverkort, J. W., Zeman, M. & Isabella, O. (2023, August). Dynamic operation of water electrolyzers: A review for applications in photovoltaic systems integration. In *Renewable and Sustainable Energy Reviews* (Vol. 182 p. 113407). Elsevier.

[6] Nedstack. "Product Data Sheet. NEDSTACK FCS 13-XXL PEM FUEL CELL STACK", Version: November 2019. https://nedstack.com/sites/default/files/2019-11/20191105_nedstack_fcs_13-xxl.pdf

[7] Chabaud, V. "FLAggTurb - Farm-Level Aggregated Turbulence", SINTEF GitLab, 2025, https://gitlab.sintef.no/valentin.chabaud/flaggturb.git

[8] Panjwani, B., Kvittem, M., Eliassen, L., Ormberg, H., & Godvik, M. (2019, October). Effect of wake meandering on aeroelastic response of a wind turbine placed in a park. In *Journal of Physics: Conference Series* (Vol. 1356, No. 1, p. 012039). IOP Publishing.

[9] Chabaud, V. B. (2024). Power Fluctuations caused by Wind Turbulence on a 15MW Floating Offshore Wind Turbine. *SINTEF Energi. Prosjektnotat*.

[10] IEA (2019, November), *Offshore Wind Outlook 2019*, IEA, Paris https://www.iea.org/reports/offshore-wind-outlook-2019

## VIII. VITA

**Maiken Omtveit** graduated with a M.Sc. degree in Environmental and Electric Engineering from the Norwegian University of Science and Technology (NTNU) in 2022. She has been a R&D Engineer with ABB AS in Oslo, Norway since 2022, and is currently doing an Industrial Ph.D. with ABB and NTNU on coordinated control between industrial process and power systems with flexible loads.

**Qian Long** received the Ph.D. degree in electrical engineering from North Carolina State University in 2020. From 2020 to 2022, he was a Postdoctoral Researcher with the Technical University of Denmark. He is currently a Senior Power Design Engineer with ABB AS in Oslo, Norway. His interests include modelling, operation, control and optimization of hybrid energy systems.

**Valentin Chabaud** received a Ph.D. in mechatronics applied to wave tank testing of floating wind turbines from NTNU in 2016. He is currently a researcher at SINTEF Energy Research with main field multidisciplinary wind farm simulation and control, focusing on system-level applications such as grid integration and asset management.

**Marte Ruud-Olsen** graduated with a M.Sc. in Cybernetics and Robotics from NTNU in 2023. She joined ABB AS in Oslo later the same year as part of their graduate program, where participants spend six months in four different ABB departments before moving on to a permanent position. Marte started in the Automation & Control team, before moving to R&D and then Service. She is currently stationed in the Czech Republic at ABB's Operation Center in Ostrava.

**Steinar Halsne** graduated with a M.Sc. degree in Environmental and Electric Engineering from the Norwegian University of Science and Technology (NTNU) in 2019. He is currently a Senior Technical Consultant with ABB as in Oslo, Norway, focusing on renewable power integrations and hydrogen systems, as well as design, modeling and control of such systems.

**Tor-Christian Ystgaard** holds a M.Sc. in Industrial Economics and Technology Management from NTNU (2005). Tor-Christian is currently leading Energy Transition Studies at ABB. Previously, he has held roles in technical consulting, engineering processes, and large electrification projects. Before ABB, he worked as Specialist Engineer at Aker Solutions, holding various systems engineering and engineering management roles.